\documentclass[review,fleqn,sort&compress]{elsarticle}
\usepackage[left=3cm,top=2cm,bottom=2cm,right=3cm,
nohead,nofoot]{geometry}

 \usepackage{graphicx}
 \usepackage[CJKbookmarks=true,bookmarksnumbered=true]{hyperref}

\usepackage{amssymb}
\usepackage{amsmath}
\usepackage{esvect}
\usepackage{graphicx}
\usepackage[caption=false]{subfig}

\usepackage{amsfonts}
\usepackage{xcolor}
\usepackage{epsfig}
\usepackage{bm}
\usepackage{tabularx}
\usepackage{multirow}
\usepackage{mathtools}
\usepackage{xfrac}
\usepackage{blkarray}
\usepackage{bbold}
\usepackage[mathscr]{euscript}
\usepackage{multimedia}
\usepackage{amsmath}
\usepackage{comment}
\usepackage{color}
\usepackage{enumitem}
\usepackage{pdfpages}
\usepackage{here}

\graphicspath{ {./Figs/} }

\title{\textbf{ Achieving Empirical Potential Efficiency with DFT Accuracy: A Neuroevolution Potential for the $\alpha$-Fe–C–H System}}

\author[adr1]{Fan-Shun~Meng\corref{cor1}\fnref{equal}}
\ead{fanshun.meng@tsme.me.es.osaka-u.ac.jp}

\author[adr1]{Shuhei~Shinzato\fnref{equal}}
\ead{shuhei.shinzato@tsme.me.es.osaka-u.ac.jp}

\author[adr1]{Zhiqiang Zhao}
\ead{zhiqiang.zhao@tsme.me.es.osaka-u.ac.jp}

\author[adr1]{Jun-Ping~Du}
\ead{jpdu@tsme.me.es.osaka-u.ac.jp}

\author[adr3]{Lei Gao}
\ead{leigao@ustb.edu.cn}

\author[adr4]{Zheyong Fan}
\ead{brucenju@gmail.com}

\author[adr1]{Shigenobu~Ogata\corref{cor1}}
\ead{ogata@me.es.osaka-u.ac.jp}

\cortext[cor1]{Corresponding authors.
Email: fanshun.meng@tsme.me.es.osaka-u.ac.jp (F.-S.Meng), ogata@me.es.osaka-u.ac.jp (S.Ogata). \fntext[equal]{These two authors contributed equally to this work.}
}

\address[adr1]{Department of Mechanical Science and Bioengineering, Graduate School of Engineering Science, Osaka University, 1-3 Machikaneyama, Toyonaka, Osaka, Japan 560-8531}
\address[adr3]{Corrosion and Protection Center , Institute for Advanced Materials and Technology, University of Science and Technology Beijing, Beijing, China 100083}
\address[adr4]{College of Physical Science and Technology, Bohai University, Jinzhou, China 121013}

\begin{document}

\begin{abstract}
A neuroevolution potential (NEP) for the ternary $\alpha$-Fe–C–H system was developed based on a database generated from spin-polarized density functional theory (DFT) calculations, achieving empirical potential efficiency with DFT accuracy. At the same power consumption, simulation speeds using NEP are comparable to, or even faster than, those with bond order potentials. The NEP achieves DFT-level accuracy across a wide range of scenarios commonly encountered in studies of $\alpha$-Fe and $\alpha$-Fe–C under hydrogen environments.
The NEP enables large-scale atomistic simulations with DFT-level accuracy at the cost of empirical potentials, offering a practical tool to study hydrogen embrittlement in steel. 
\end{abstract}


\maketitle
\newpage
\section{Introduction}
\label{sec:intro}

Hydrogen embrittlement (HE), first reported in 1875 \cite{johnson1875some}, remains a significant challenge for steels and other metals, as it undesirably reduces their ductility and fracture toughness, representing a critical scientific challenge in building reliable infrastucture for a carbon-free hydrogen economy\cite{li2025synergistic}. Elucidating the mechanisms of HE requires a comprehensive investigation of the structural evolution of steel in hydrogen-containing environments. Achieving this goal necessitates a robust atomistic computational framework that not only captures the interactions among hydrogen, carbon, and the iron matrix with high fidelity, but also scales to large atomistic models to compute diverse Fe defects and their coupled interactions. However, direct first-principles modeling of large scale steel–hydrogen systems remains computationally difficult due to the substantial cost associated with density functional theory (DFT) calculations. To address this limitation, several empirical interatomic potentials have been developed for the ternary Fe--C--H systems \cite{zhou2020review,mun2021modified,islam2016interactions}, allowing large scale simulations. Nonetheless, accurately modeling systems containing multiple defects remains challenging, as empirical potentials inherently lack the transferability and fidelity required to describe complex defect interactions.

Benefiting from the high flexibility of neural networks, several types of machine learning interatomic potentials(MLIPs) have been proposed\cite{behler2007generalized,bartok2010gaussian,drautz2019atomic,novikov2020mlip,zeng2023deepmd,cheng2024cartesian}. MLIPs can capture a wide range of material properties and address long-standing challenges of empirical potentials\cite{ying2025advances}, such as the accurate description of screw dislocation in $\alpha$-Fe \cite{maresca2018screw,mori2020neural,allera2025activation}. Recently, we constructed a neural netowrk interatomic potential (NNIP) for the ternary system\cite{meng2025high} under the frame work of Behler and Parrinalo\cite{behler2007generalized} using the n2p2 package\cite*{singraber2019parallel}. The NNIP allows simulations of systems containing hundreds  of thousands of atoms at the DFT accurate and nanosecond scale; however, for systems comprising millions of atoms, the required computational resources (in core-hours) often exceed practical limits.

The Neuroevolution Potential (NEP) \cite{fan2021neuroevolution}, a type of NNIP, utilizes the atomic cluster expansion approach \cite{drautz2019atomic} as its descriptor for local atomic environments, incorporates a mixed-precision strategy to balance accuracy and efficiency, and can be directly used to perform molecular dynamics simulations on graphics processing units (GPUs). Crucially, all per-atom quantities have closed-form expressions, allowing a one-to-one mapping between each atom and a CUDA (compute unified device architecture) thread. This parallelization scheme proves highly efficient for medium- to large-scale systems, as it achieves both a high level of parallelism and substantial arithmetic intensity, satisfying the two critical factors for maximizing GPU performance.\cite{fan2021neuroevolution}. All above factors results in a exceptionally high efficiency of NEP on GPUs.

 In this work, we constructed a NEP for the  ternary $\alpha$-Fe–C–H system using the database prepared for the NNIP training in Ref.\cite{meng2025high}. This NEP attains accuracy comparable to that of the NNIP (at the DFT-level), while achieving a computational cost comparable to, or even lower than, that of bond-order potentials (BOPs).  The NEP with GPUs enables large-scale atomistic simulations with DFT-level accuracy at the cost of empirical potentials, offering a practical tool to study hydrogen embrittlement in steel.
 
\section {Methodology}

The NEP model, whose latest version is NEP4, is generally under the Behler-Parinello framework with different atomic-environment descriptors (AEDs) and the training method. Two types of the AEDs, radial and angular descriptors,  were employed. 
The radial descriptors of atom $i$ are labeled by the index $n$ and are constructed as a sum of radial functions over the neighboring atoms $j$:
\begin{equation}
q^i_n=\sum_{j(j\neq i)}g_n(r_{ij}) ~~~ \text{with}~~~~~ 0 \leq n \leq n^{\text{R}}_{\text{max}}.
\label{rd}
\end{equation}
The radial funciton $g_n(r_{ij})$ is constructed as a linear combination of  $N^{\text{R}}_{\text{bas}}+1$ basis functions $f_k(r_{rj})$:
\begin{equation}
g_n(r_{ij})=\sum_{k=0}^{N^\text{R}_\text{bas}}c_{nk}^{IJ}f_k(r_{rj})
\label{bf}
\end{equation}
in which $N^{\text{R}}_{\text{bas}}$ is the number of radial descriptors, and the basis functions $f_k(r_{rj})$ are defined as:
\begin{equation}
f_k(r_{rj})=\frac{1}{2}[T_k(2(r_{ij}/r^R_c-1)^2-1)+1]f_c(r_{ij})
\label{fk}
\end{equation}
where $T_k(x)$ is the $k$-th order Chebyshev polynomial of the first kind. $f_c(r_{ij})$ is the cutoff function and defined as
\begin{equation}
f_c(r_{ij})=
\begin{cases}
\frac{1}{2}[1+\text{cos}\left(\pi\frac{r_{ij}}{r_c^R}\right)] &   r_{ij} \le r_{c}^R; \\
0 &  r_{ij} > r_{c}^R.
\end{cases}
\label{cf}
\end{equation}
In Equ. \ref{rd} to \ref{cf}, $r_{ij}$ indicates the distance between atom $i$ and its neighboring atom $j$. $r_\text{c}^\text{R}$ is the cutoff distance, $c_{nk}^{IJ}$ are the expansion coefficients depending on $n$ and $k$ and also on the types of atoms $i$ and $j$, which are parameters to be trained. Beyond $N_\text{max}^\text{R}$, $N^\text{R}_\text{bas}$ and $r_\text{c}^\text{R}$ are all hyperparameters. 

The angular descriptors of the center atom $i$ depend on both radial distances $r_{ij}$ and the angles $\theta_{ijk}$ enclosed by $\mathbf{r_{ij}}$ and $\mathbf{r_{ik}}$.
The typical angular descriptors in NEP4 are defined in terms of Legendre polynomials $P_l(x)$:
\begin{equation}
q_{nl}^i=\frac{2l+1}{4\pi}\sum_{j\neq i}\sum_{k\neq i}g_n(r_{ij})g_n(r_{ik})P_l(\text{cos}\theta_{ijk}), ~~\text{where}~~ \text{cos}_\text{ijk}=\frac{\mathbf{r_{ij}}\cdot \mathbf{r_{ik}}}{r_{ij}~ r_{ik}}.
\label{ang}
\end{equation}
The radial and angular dependencies are indicated by the subscripts $n$ and $l$ in $q_{nl}^i$. Note that the radial functions in $q_{nl}^i$ are defined similarly to Equ.\ref{bf} but employing a different cutoff radius $r_\text{c}^\text{A}$ and expansion order $N_\text{bas}^\text{A}$. There are also other types of angular descriptors in NEP, please refer to Ref.\cite{fan2022gpumd}

Using the atomic configurations in the training database and the above two types of structure descriptors, one can get the vector of $\text{q}^i$ as the input of the neural network, this information will transfer through the hidden layer via the active function (tanh(x)) and bias for each neuron ($b_\mu^{(0)}$) to the output layer $U_i$ ($b^{(1)}$ is the bias of $U_i$), i.e. the atomic energy of the atom $i$. 
\begin{equation}
U_i=\sum_{\mu=1}^{N_{neu}} w_\mu^{{1}} \text{tanh}\left( \sum_{v=1}^{N_{\text{des}}}w_{\mu v}^{(0)}q_v^i-b_\mu^{(0)}\right)-b^{(1)}
\label{ae}
\end{equation}
All wights $w$ and bisas $b$ are trainble parameters.

The NEP for the $\alpha$-Fe–C–H ternary system (hereafter referred to as the NEP) was trained using the open-source GPUMD-4.2 package\cite{xu2025gpumd,gpumdgithub} with the NEP4 model\cite{xu2025gpumd}. We considered up to 4-body interactions in the atomic environment descriptors\cite{fan2022gpumd}, all elements shared the same cutoff radii for the radial and angular descriptors, which are 6.0 and 5.0 \AA, respectively. The Chebyshev polynomial expansion orders are set to 8. The number of neurons in the single hidden layer of the neural network is 100.   
The population size was set to 100, and the training was executed with a batch size of 20000 structures using the separable natural evolution strategy (SNES)\cite{schaul2011high}. The loss function incorporated contributions from energy and forces, whereas the contribution from virial stress was excluded due to the absence of the corresponding information in the database. Additionally, a regularization term ($L_{2}$) was included in loss function to mitigate the risk of overfitting:
\begin{equation}
L_{2}(\mathbf{z}) = \left( \frac{1}{N_{\text{par}}} \sum_{n=1}^{N_{\text{par}}} z_{n}^{2} \right)^{1/2},
\label{lmda2}	
\end{equation}
where $z$ and $N_{\text{par}}$ are the trainable parameters and the total number of these parameters, respectively\cite{fan2021neuroevolution}.  The weights assigned to the energy, force, and regularization contributions in the loss function were $1.0$, $1.0$, and $0.05$, respectively.

\begin{figure}[H]
    \centering
\includegraphics[width=0.64\textwidth]{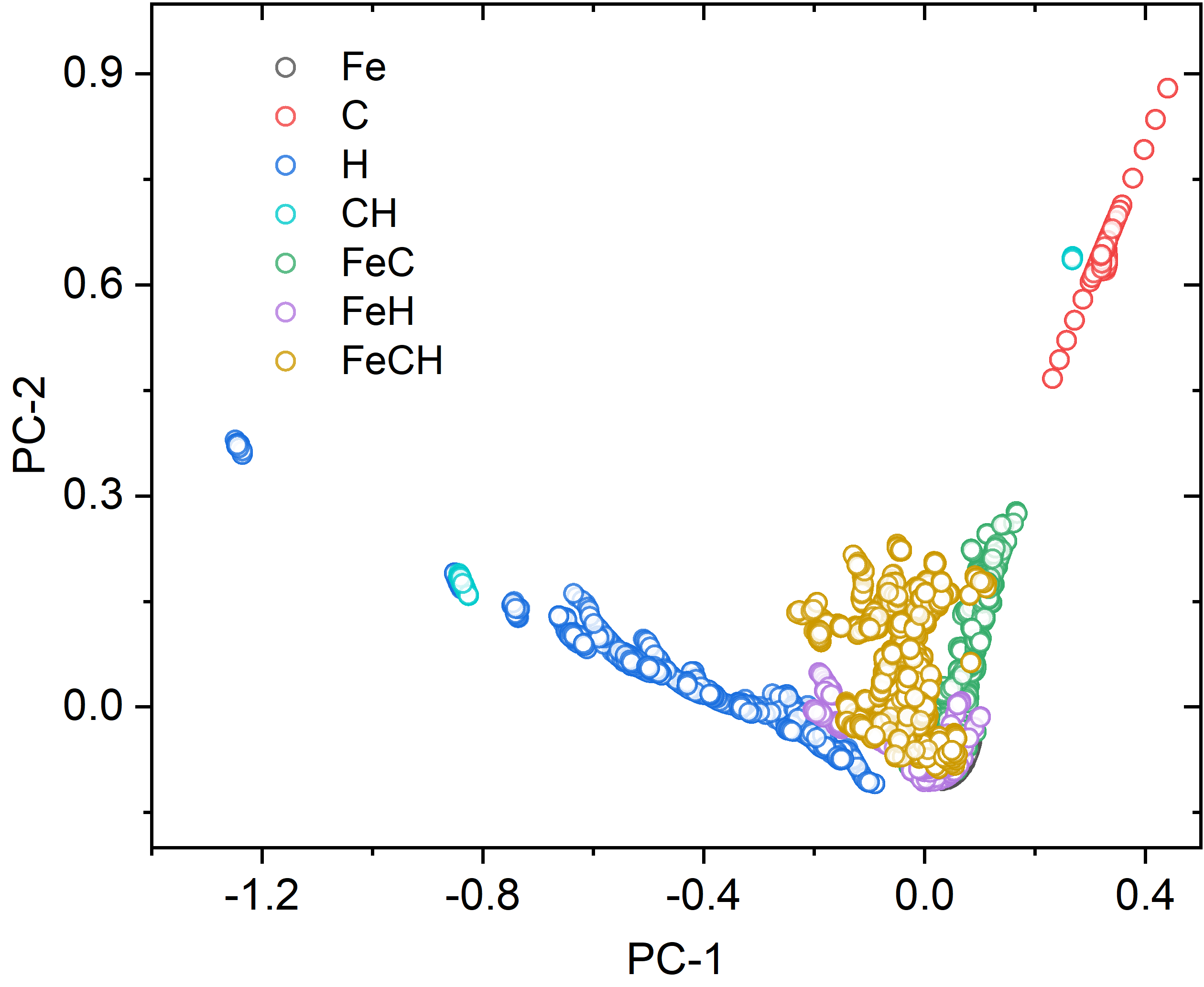}
\caption{\label{pca} Visualization of the database in the descriptor space using the Principal Component Analysis. PC indicates the principal component.}
\end{figure}

The dataset was adopted from our previous work\cite{meng2025high}, which has $62743$ configurations, equivalent to $6.05\times 10^{6}$ local atomic environments.  Following the method proposed in Ref.\cite{zhao2024general}, the database was visualized in the descriptor space using the approach of Principal Component Analysis \cite{abdi2010principal}. As the result shown in Fig. \ref{pca}, the database includes various datasets for unary, binary, and ternary systems, which can cover a wide range of local atomic environments. Detailed information can be found in the supplementary materials of Ref.\cite{meng2025high}. Ten percent of the data were randomly selected for testing, while the remaining data were used for the NEP training. 

\section{Efficiency of the NEP for the $\alpha$-Fe--C--H system}
To evaluate the computational efficiency and power consumption of the NEP relative to the NNIP\cite{meng2025high} and BOP\cite{zhou2020review}, we systematically performed identical simulations on both CPU and GPU platforms. For the CPU platform, a single node equipped with two Intel Xeon Platinum 8360Y CPUs (36 cores/CPU$\times$2 CPUs=72 cores) was used, and the NEP, NNIP, and BOP were all tested on this platform. Tests using GPU were conducted separately on the NVIDIA A100-SXM4-40GB and H100-80GB HBM platforms, with only a single GPU per node utilized, despite each node being equipped with eight GPUs; Only the NEP was tested on these GPU platforms because NNIP and BOP currently do not have good support for GPU acceleration. The three platforms, referred to here as CPU, A100, and H100, have corresponding power consumptions of 500 W, 400 W, and 700 W, respectively.

\begin{figure}[H]
    \centering
\includegraphics[width=0.64\textwidth]{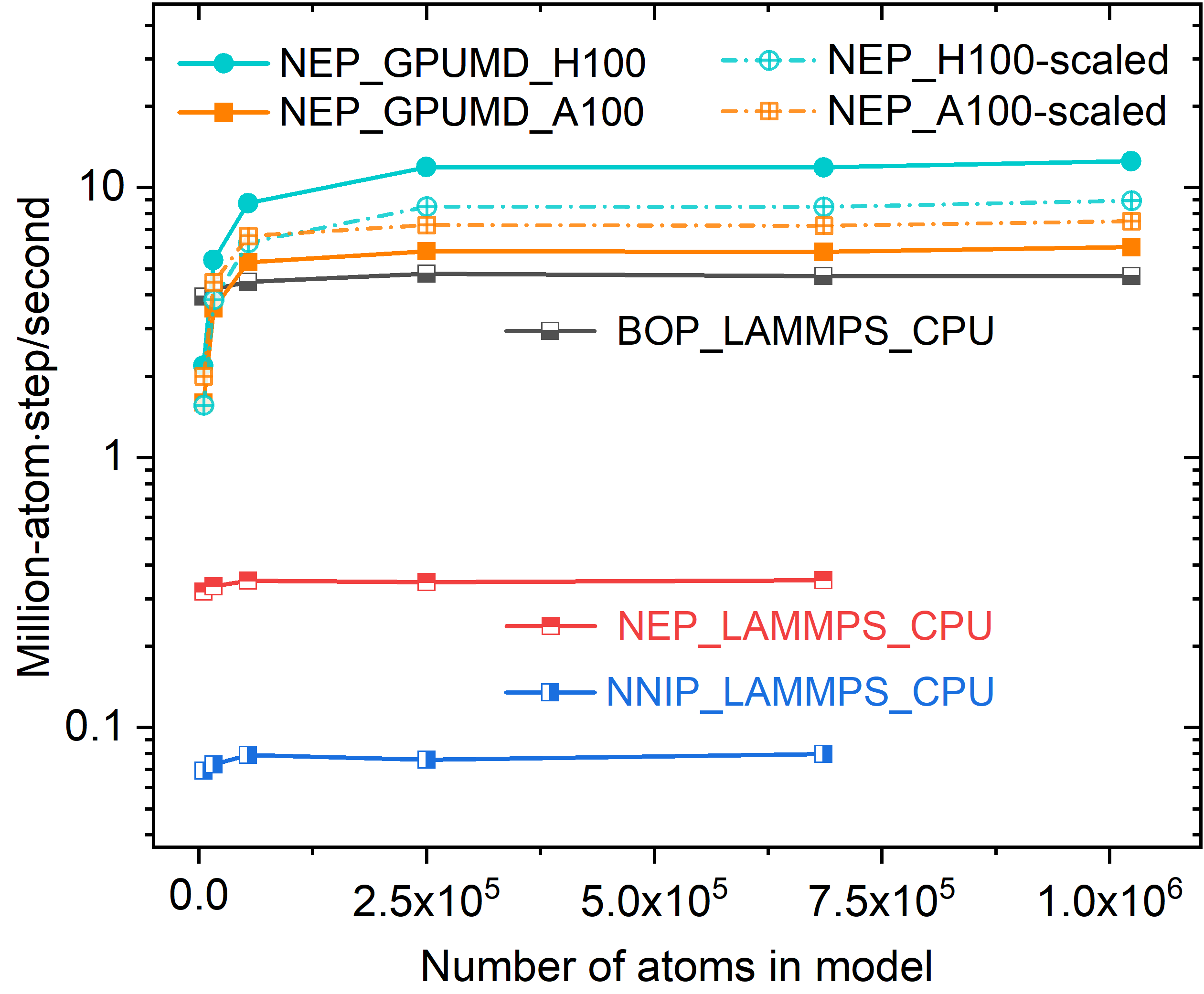}
\caption{\label{cost} Computational cost of NEP compared with that of  NNIP and BOP using LAMMPS on CPU and using GPUMD on GPU platforms. NEP${\_}$ H/A100-scaled indicates that the computational costs were normalized with respect to the power consumption (in watts) of the CPU and GPU employed in this test, see the main text.}
\end{figure}

Five models, labeled model-I to model-VI, were created with $5,070$, $16,000$, $54,000$, $250,000$, $686,000$, $1,024,000$ Fe atoms in BCC lattice, into which $0.16~\rm{at.}\%$ C and $1.0~\rm{at.}\%$ H atoms were randomly introduced. The LAMMPS\cite{plimpton1995fast}  and GPUMD\cite{fan2022gpumd,gpumdgithub} packages were used for the tests on the CPU and GPU platforms, respectively. The test results are summarized in Fig.~\ref{cost}.

On the CPU platform, the efficiency of the NNIP is approximately 0.07 million-atom$\cdot$step/second for models containing no more than 20,000 atoms. Remarkably, NEP achieves a performance of 0.32 million-atom$\cdot$step/second, which is about four to five times faster than NNIP, and this efficiency advantage is maintained as the model size increases. On this platform, BOP exhibits the highest efficiency, reaching 4.2 million-atom$\cdot$step/second, which is about 60 times faster than NNIP and 13 times faster than NEP. For medium- to large-sized models (Model III to Model V), the speedup factors of BOP are 13.4 relative to NEP and 58.9 relative to NNIP, respectively.

The efficiency of NEP was assessed on the two GPU platforms. For small-sized models Model I (and Model II, respectively) the efficiency increased from 0.32 (0.33) million-atom$\cdot$step/second on the CPU platform to 1.60 (3.57) and 2.19 (5.38) million-atom$\cdot$step/second on the A100 and H100 platforms, respectively. The GPU resources are fully utilized when the model size exceeds 54,000 atoms on the A100 platform and 250,000 atoms on the H100 platform, achieving efficiencies of 5.3 and 11.9 million-atom$\cdot$step/second, respectively, showing 69.7 and 156.6 times faster than that of the NNIP,  1.1 and 2.5 times faster than that of the BOP, on the CPU platform.

Considering power consumption, the computational efficiency of the NEP on the GPU platforms was normalized with respect to the CPU power by multiplying a factor of $f=\frac{W_{\rm{cpu}}}{W_{\rm{gpu}}}$, where the $W_{\rm{gpu}}$ and $W_{\rm{cpu}}$ are the power of GPU and CPU respectively. The value of $f$ is 1.25 for A100 and 0.714 for H100, respectively. As shown by the dotted line in Fig.~\ref{cost}, the NEP efficiency for model-V is 7.22, and 8.49 million-atom$\cdot$step/second on A100 and H100, while that of the BOP on CPU is 4.70 million-atom$\cdot$step/second,  indicating the efficiency of NEP on A100 and H100 is higher (also power consumption is lower) than that of the BOP on CPU. 

To demonstrate the practical computational efficiency of the NEP developed in this study using the GPUMD package, a large-scale MD simulation was performed for a system comprising over 5 million atoms (5,488,000 Fe, 9,144 C, and 54,888 H) with a simulation box size of 404.5$\times$404.5$\times$404.5 Å$^3$, adopting periodic boundary conditions in all directions. The simulation was carried out in the NVT ensemble at 300 K with a time step of 0.5 fs, utilizing eight H100 GPUs on a single node. A 10 ps simulation was completed in 1,817 seconds. The computational speed reached 61.1 million-atom$\cdot$step/second, indicating that about 50 hours would be sufficient to perform a 1 nanosecond MD simulation for a system of comparable size and time step.
 
\section{Accuracy of the NEP for the $\alpha$-Fe--C--H system}
\subsection{Overall accuracy of the $\mathrm{NEP}$}
\label{subsec:accur}

\begin{figure}[H]
    \centering
\subfloat[(a)]{\includegraphics[width=0.48\textwidth]{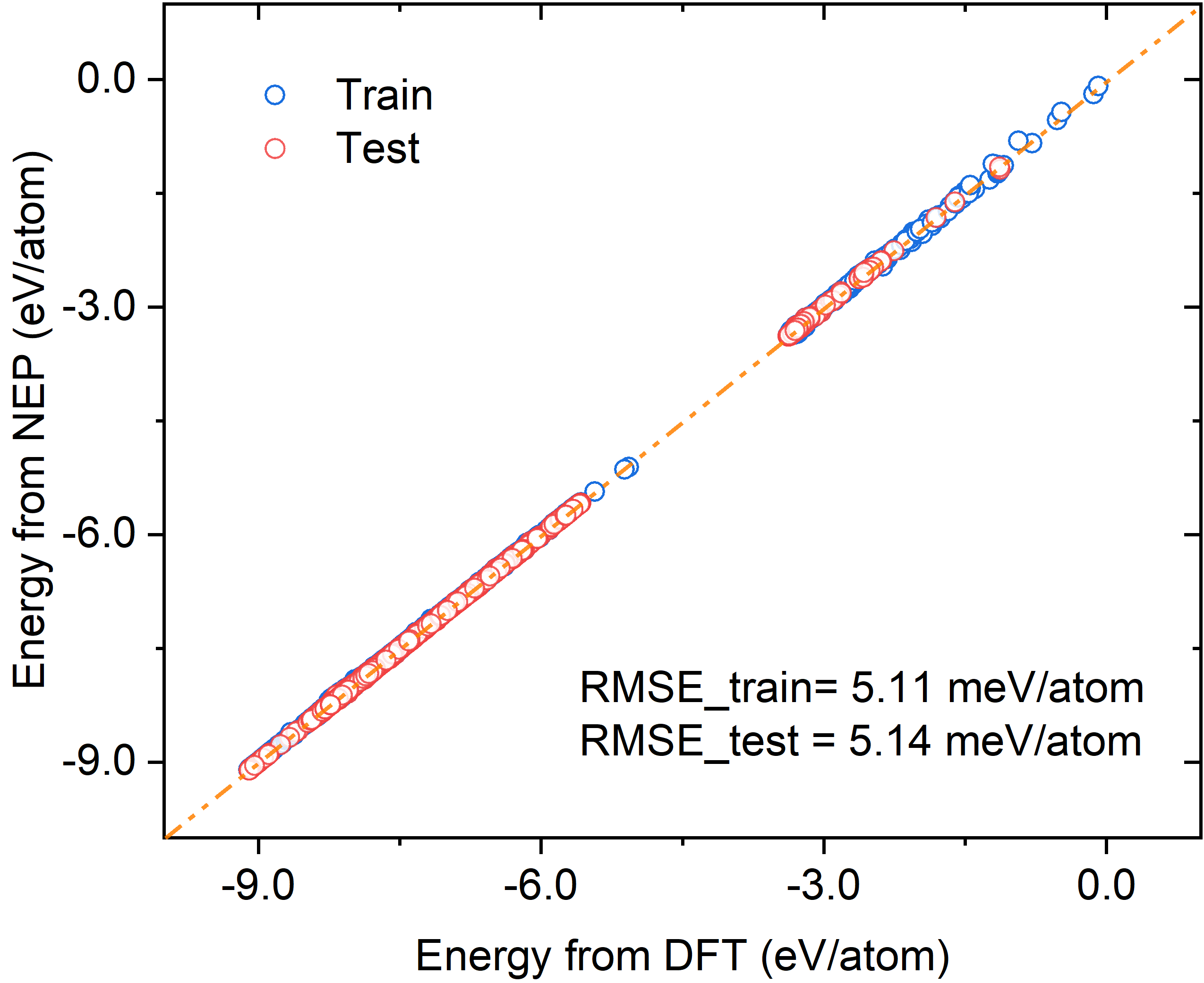}}
\hfill
\subfloat[(b)]{\includegraphics[width=0.48\textwidth]{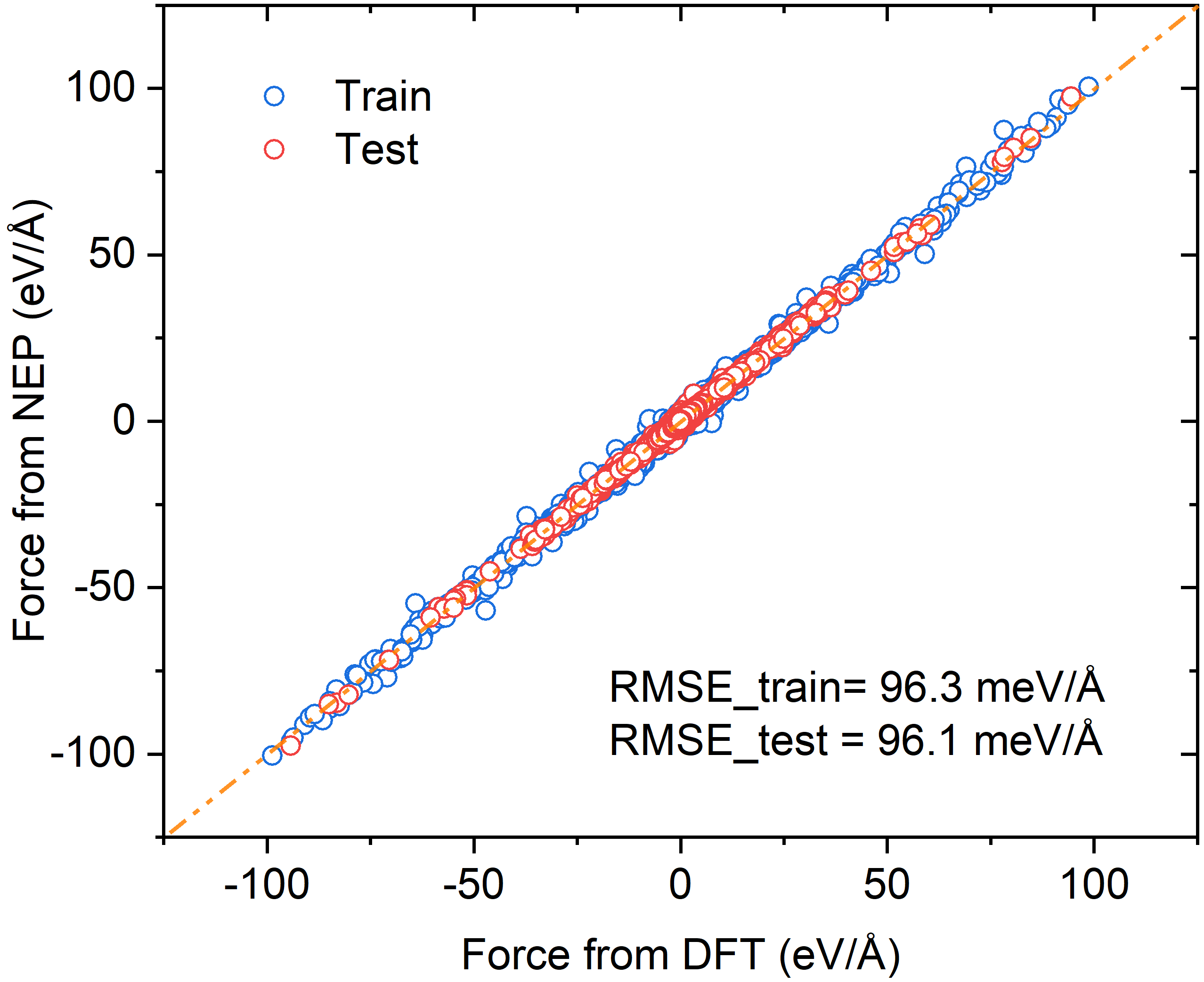}}
\caption{\label{Fig1} Comparison of NEP and DFT (a) energies and (b) forces of the structures in the training and testing data sets. The dotted line with a slope of 1 corresponds to a perfect training.}
\end{figure}

The overall accuracy of the NEP was indicated using the root mean squared error (RMSE) of energies ($E$) and forces ($F$). The training and testing data sets have ${\rm RMSE}(E)$s of $5.11$ and $5.14$ meV/atom, and ${\rm RMSE}(F)$s of $96.3$ and $96.1$ meV/\AA, respectively, indicating the trained NEP exhibits comparable accuracy with DFT calculations. The values of these RMSEs are slightly higher than those from the NNIP\cite{meng2025high}, but remain within the typical range for machine learning based interatomic potentials\cite{zhao2023development}.  
The point by point comparison of DFT and NEP produced energies and atomic forces of the structures in the training and testing data sets are supplied in Fig.~\ref{Fig1}. All of those points of the training and testing datasets are distributed along the line with a slope of 1, indicating that the training dataset can be well reproduced and no overfitting occurred.

\subsection{The NEP performance for the fundamental properties of $\alpha$-\rm{Fe-C-H}}
\label{nep-bop-nnip}
The performance of the NEP for the fundamental properties of the $\alpha$-Fe-C-H system, in comparison with those of the NNIP\cite{meng2025high}, BOP\cite{zhou2020review}, is summarized in Table. \ref{table1}. Note that Ref. \cite{zhou2020review} reported two BOPs with comparable performance, and the results from BOP-I were adopted for the comparison. Additionally, more information regarding the NEP performance is provided in the supplementary materials.

All properties listed in Table.\ref{table1} from the NEP show good agreement with those from the NNIP and our own DFT calculations, with the exception of the elastic constant $C_\text{12}$, which is $23\%$ lower than the reference value. It is unclear why the DFT results for $E_\text{c}$ and $E_\text{C/H@O/T}$ reported in Ref. \cite{zhou2020review} differ significantly from our own DFT results. The results from BOP are less accurate than those from the two MLIPs, due to the well-known limitation of empirical potentials arising from their fixed formalism.

\begin{table}[H]
\centering
\caption{\label{table1} Comparison of the fundamental properties of the ternary system from the NEP, NNIP, BOP, and DFT. $E_\text{c}$ is the cohesive energy of Fe atom with equivalent lattice constant in bcc lattice.  $E_\text{C@O}$ and $\Omega_\text{C@O}$ denote the formation energy and the corresponding volume expansion of an interstitial C atom at an O site in bcc iron, respectively. Note that, $E_\text{C@O}$ is the energy difference between a $4\times4\times4$ bcc Fe model  with and without a C atom at O-site, i.e., $E_\text{C@O}$=$E_\text{(128Fe+1C@O)}$-$E_\text{128Fe}$. The same notation applies to other cases. $E_\text{H@vac}$ is the formation energy difference of H atom in bcc Fe model and in a vacancy, i.e., $E_\text{H@vac}$=$E_\text{(128Fe+1H@T)}$-$E_\text{(127Fe+1H@Vac)}$. The BOP-I in Ref.\cite{zhou2020review} was adopted in this work.\\}

\resizebox{13.6cm}{!}{%
\begin{tabular}{lccccc}
\hline
Property & NEP & NNIP\cite{meng2025high} & BOP\cite{zhou2020review} & DFT-This work & DFT\cite{zhou2020review} \\
\hline
Lattice constant, $a_0$, \AA       & 2.833  & 2.829  & 2.889  & 2.830  & 2.830 \\
Cohesive energy, $E_c$, eV/atom     & -8.242 & -8.241 & -4.180 & -8.241 & -5.215 \\
Elastic constant, GPa \\
$C_{11}$  & 267    & 283    & 218    & 297    & 272 \\
$C_{12}$  & 116    & 145    & 142    & 151    & 155 \\
$C_{44}$  & 103    & 91     & 130    & 105    & 103 \\
Formation energy, eV \\
$E_\text{C@O}$     & -8.281 & -8.402 & -6.190 & -8.414  & -5.252 \\
$E_\text{C@T}$     & -7.497 & -7.606 & -5.870 &   --   & -5.236 \\
$E_\text{H@O}$    & -3.006 & -3.019 & -2.170 &   --   & -5.156 \\
$E_\text{H@T}$     & -3.148 & -3.164 & -2.270 & -3.218 & -5.159 \\
Volume expansion, \AA$^3$ \\
$\Omega_\text{C@O}$     &  9.798 & 10.829 &  4.200 &   --   & 11.110 \\
$\Omega_\text{C@T}$     & 10.153 & 10.299 &  7.370 &   --   & 10.320 \\
$\Omega_\text{H@O}$    &  3.660 &  4.588 &  1.830 &   --   &  4.220 \\
$\Omega_\text{H@T}$     &  3.688 &  4.544 &  4.480 &   --   &  4.230 \\
Trapping energy of H in vacancy, eV \\
$E_\text{H@vac}$     & -0.602 & -0.598 & -0.080 & -0.592 &   --   \\
(110) Surface energy, $\gamma_\text{110}$, J/m$^2$ & 2.375 & 2.438 & 1.344 & 2.449 & 2.370 \\
\hline
\end{tabular}
}
\end{table}

The BOP was constructed based on the Fe–C BOP developed for carbides \cite{henriksson2009simulations}. The lattice constants, elastic constants for the typical carbides of cementite ($\text{Fe$_3$C}$) from the NEP, NNIP, BOPs, and available DFT as well as experimental results are tabulated in Table.\ref{table2}.  The results obtained from the NEP are generally similar to those from NNIP and DFT; however, $C_\text{44}$ obtained from BOP is significantly overestimated.  The phonon dispersion curves of $\text{Fe$_3$C}$ from the NEP, BOP are provided in the supplementary materials, aligning well with those from the NNIP and DFT.
\begin{table}[H]
\centering
\caption{\label{table2} Lattice constants (\AA) and elastic constants (GPa) from the NEP, BOP, NNIP, DFT and experiment.}
\resizebox{13.0cm}{!}{%
\begin{tabular}{lcccccc}
\hline
Property  & NEP & NNIP\cite{meng2025high}  & BOP\cite{zhou2020review} & BOP \cite{henriksson2009simulations} & DFT\cite{jiang2008structural,nikolussi2008extreme} & Exp\cite{wood2004thermal} \\
\hline
Lattice constant, \AA \\
$a$ & 5.058 & 5.023 & 4.961 & 5.086 & 5.04 & 5.036 \\
$b$ & 6.710 & 6.747 & 6.473 & 6.521 & 6.72 & 6.724 \\
$c$ & 4.460 & 4.461 & 4.477 & 4.498 & 4.48 & 4.480 \\
Elastic constants, GPa \\
$C_\text{11}$ & 379 & 371 & 332 & 363 & 388 & - \\
$C_\text{22}$ & 286 & 358 & 364 & 406 & 345 & - \\
$C_\text{33}$ & 279 & 333 & 377 & 388 & 322 & - \\
$C_\text{12}$ & 153 & 186 & 184 & 181 & 156 & - \\
$C_\text{13}$ & 146 & 155 & 170 & 166 & 164 & - \\
$C_\text{23}$ & 126 & 202 & 134 & 130 & 162 & - \\
$C_\text{44}$ & 24 & 15 & 69 & 91 & 15 & - \\
$C_\text{55}$ & 123 & 131 & 118 & 125 & 134 & - \\
$C_\text{66}$ & 119 & 116 & 127 & 134 & 134 & - \\
\hline
\end{tabular}
}
\end{table}

\begin{figure}[H]
    \centering
\includegraphics[width=0.78\textwidth]{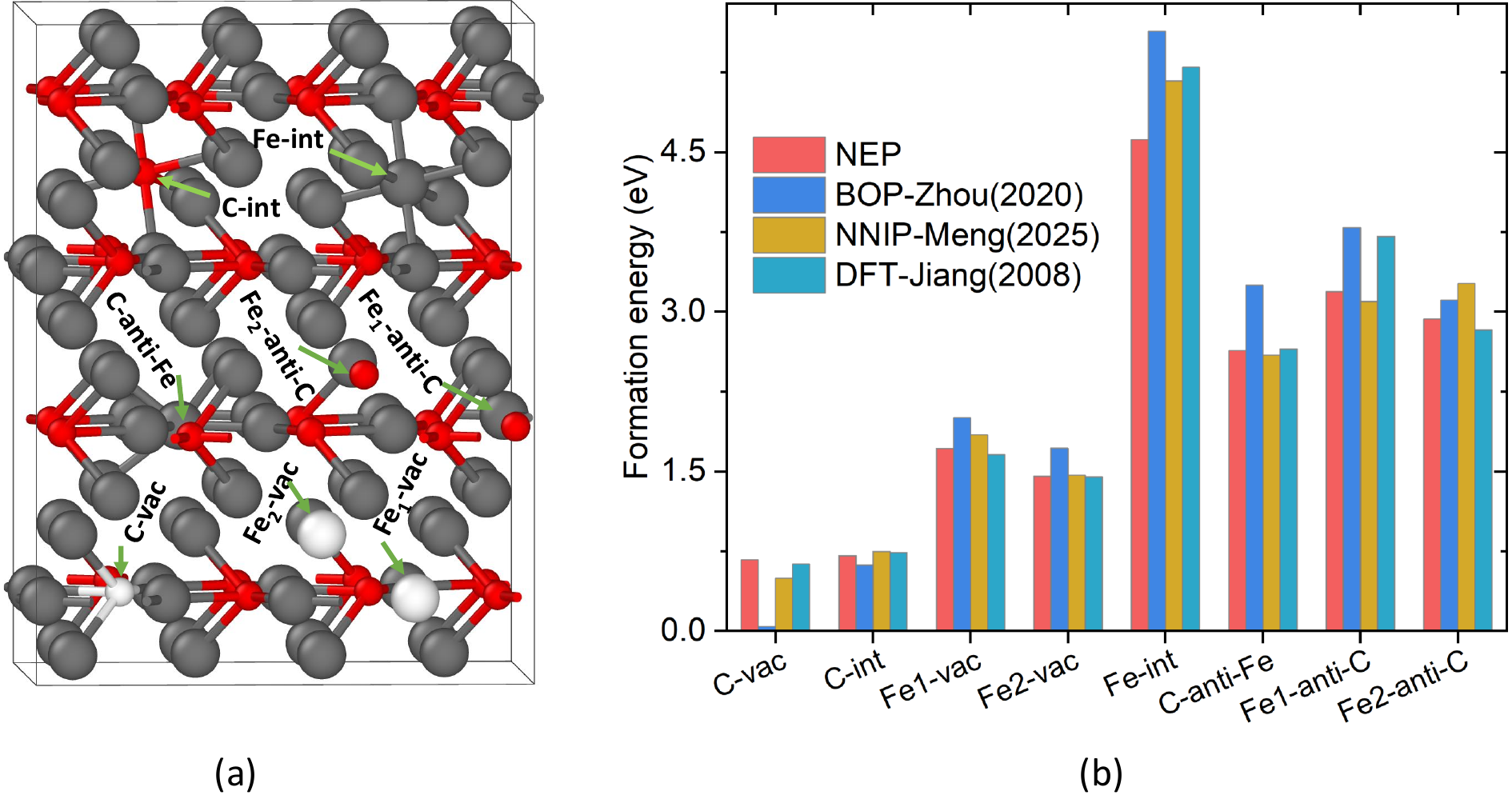}
\caption{\label{def-Fe3C} Point defects in Fe$_3$C (a) and the corresponding formation energies (b).  C-vac, C-int, Fe$_1$-anti-C in (a) stand for C vacancy, interstitial atom of C, and Fe$_1$ atom replaced by a C atom, same rule can be applied to other cases. The defect formation energy from BOP\cite{zhou2020review}, NNIP\cite{meng2025high} and DFT\cite{jiang2008point} were also plotted in (b).}
\end{figure}

Further more, the formation energy of 8 types of point defects in Fe$_3$C  were studied using NEP, and compared with those from BOP, NNIP, and DFT. The defects are depicted in Fig. \ref{def-Fe3C}(a) and the corresponding formation energy are plotted in Fig. \ref{def-Fe3C}(b).\\

Taking the results from DFT calculations reported by Jiang et al. \cite{jiang2008point} as the reference, all models yield reasonable results for the defects, except the C-vacancy formation energy predicted by the BOP, which is significantly underestimated.

\subsection{The NEP performance for C/H in screw dislocation}
\label{screw+H}
Beyond its advantages over BOP in terms of efficiency and fundamental properties addressed above, the NEP also offers several additional benefits.
Given the importance of screw dislocations in bcc metals\cite{Proville2013prediction,shinzato2019atomistically,li2025synergistic}, the performance of the NEP in this regard is presented and compared with that of NNIP and DFT calculations, while results from the BOP are excluded because it did not account for properties related to screw dislocation\cite{muller2007analytic}. 

The NEP can correctly describe the core configurations of the screw dislocation, a task that is challenging for most Fe-based empirical potentials. Relative to the easy-core configuration, the energies of the hard-core, split-core, and middle-point configurations are 27.6, 104.2, and 32.1 $\text{meV/}b$ ($b$ is the Burgers vector length), respectively.  These results are in line with DFT results of 39.3, 108, and 37.9$\text{meV/}b$\cite{itakura2013effect}, and NNIP results of 36.1, 34.2, and 108.0 $\text{meV/}b$\cite{meng2025high}, respectively. Using the NEP with the CI-NEB method\cite{mills1995reversible}, dislocation gliding follows the kink mechanism, exhibiting a barrier of 0.67 eV, showing agreement with the DFT-based line tension model of 0.73 eV\cite{itakura2013effect}, 0.86 eV\cite{Proville2013prediction}, and that from NNIP of 0.72 eV\cite{meng2025high}, 0.70 eV\cite{meng2021general}.


Recent DFT calculations indicated that H atom can stabilize the hard-core configuration of screw dislocation\cite{borges2022ab}, and C atom decoration can drive the core reconstruction occurring from the easy core to the hard core\cite{ventelon2015dislocation}, due to the interaction between H/C with dislocation. We evaluate the performance of the NEP with respect to this aspect.

To study of the interaction between H atoms and screw dislocation, models with hard-core configuration were adopted. Three H atoms were (1 H atom was) positioned at the facet (center of the prism) of the hard-core configuration with varying separations along the Burgers vector direction. The interaction energy can be determined with reference to the binding energy of a H@T-site in bcc Fe. For the 3 H case, the hard core configuration was kept up to the H atoms separation of 5$b$ during the structure relaxation. The interaction energies are  -0.507, -0.498, -0.462, -0.440, and -0.417 eV for the separations of $1b$ to $5b$, respectively, shows agreement with those from DFT calculations of -0.39, -0.35, -0.33, -0.28, and  -0.26 eV\cite{borges2022ab} and NNIP results of -0.378, -0.466, -0.399, -0.393, and -0.371 eV\cite{meng2025high}. For the single H atom case, the hard-core configuration can be maintained up to a H separation of 2$b$ with the interaction energy of -0.324 and -0.313 eV, whereas the hard core becomes distorted from the H separation of 3$b$. Configurations of H-screw dislocation are provided in the supplementary materials.

For the interaction between C atom and the screw dislocation, models with the easy-core configuration were adopted. A single C atom was initially positioned at the facet of the core of the model with dislocation length of $1b$, and the core was transformed to the hard-core configuration during the structure relaxation, with the carbon atom located at the center of the prism of the hard core. This structure transformation vividly agree with that from DFT calculations\cite{ventelon2015dislocation}. The C-screw dislocation interaction energies are -0.808, -1.022, -1.026, -0.994, -0.974, and -0.951 eV for the C-C separation of $1b$ to $6b$, respectively. These results exhibit the same trend as those obtained from NNIP and DFT calculations, being approximately 0.2 eV lower than the NNIP results and 0.22 eV lower than the DFT values, which is partly due to differences in the binding energy of C at a O-site in bcc iron.
\begin{figure}[H]
    \centering
\includegraphics[width=0.96\textwidth]{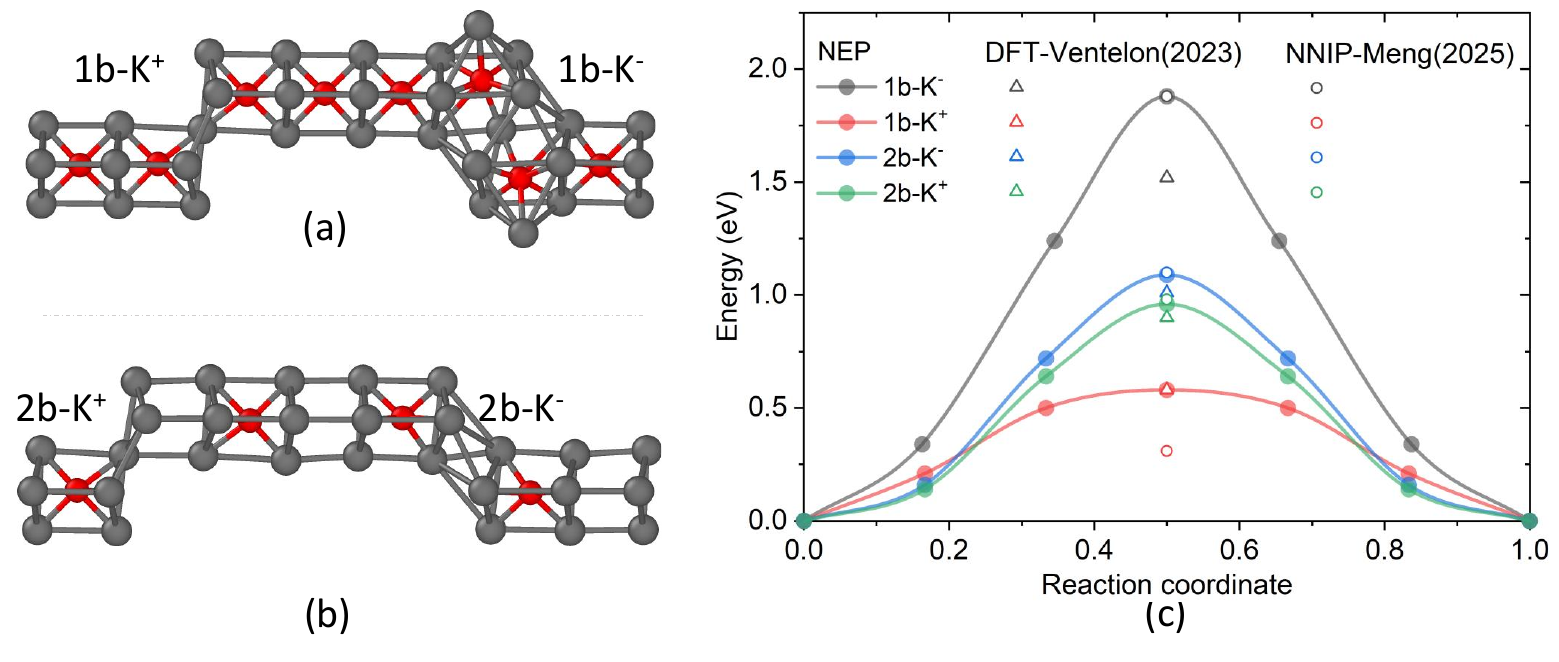}
\caption{\label{k+k-} Configurations of C-decorated kinks and their migration energy barriers. (a)-(b) local atomic configuration of Kinks with C-C separation of 1b and 2b in the Burgers vector direction. $\text{K}^+$ and $\text{K}^-$ stand for the nonequivalent kinks. Gray and red balls stand for Fe and C atoms, respectively.  (c) Energy barriers for the 4 types of C-decorated kinks, compared with results from NNIP\cite{meng2025high} and DFT\cite{ventelon2023mobility}.}
\end{figure}

The segregation of C to screw dislocations also strongly influences both the kink nucleation and migration barriers\cite{ventelon2023mobility}, which is believed to account for the strengthening effect of carbon in steels\cite{allera2022carbon}. The kink nucleation energy for a C-decorated screw dislocation with C-C separation of 2$b$ from the NEP is 1.42 eV, agrees with that from DFT calculations of 1.19 eV\cite{ventelon2023mobility} and NNIP of 1.36 eV. 
As shown in Figs.~\ref{k+k-}(a)–(b), two nonequivalent kinks, denoted as $\text{K}^+$ and $\text{K}^-$, were formed on the screw dislocations, where 1$b$ and 2$b$ in figures indicate the C–C separations along the dislocation line.  These configurations are consistent with those from NNIPs\cite{meng2025high, meng2024highly} and similar as those from DFT calculations\cite{ventelon2023mobility}.
The migration energy barriers of the four types of C-decorated kinks were further evaluated using the CI-NEB method, and the results are plotted in Figs.~\ref{k+k-}(c), together with those obtained from NNIP and DFT calculations.  The migration energy barriers predicted by the NEP for 1b-$\text{K}^+$, 1b-$\text{K}^-$, 2b-$\text{K}^+$, and 2b-$\text{K}^-$ kinks are 0.58, 1.88, 0.96, and 1.09 eV, respectively, comparable to those from DFT (NNIP) of 0.58 (0.31), 1.52(1.88), 0.90(0.98), and 1.01(1.10) eV, respectively. The ability to correctly and accurately describe screw dislocation is an advantage of the NEP over the BOP.

\subsection{The NEP performance for H in cementite and ferrite-cementite interface}
\label{H-Fe3C}
\begin{figure}[H]
    \centering
\includegraphics[width=0.72\textwidth]{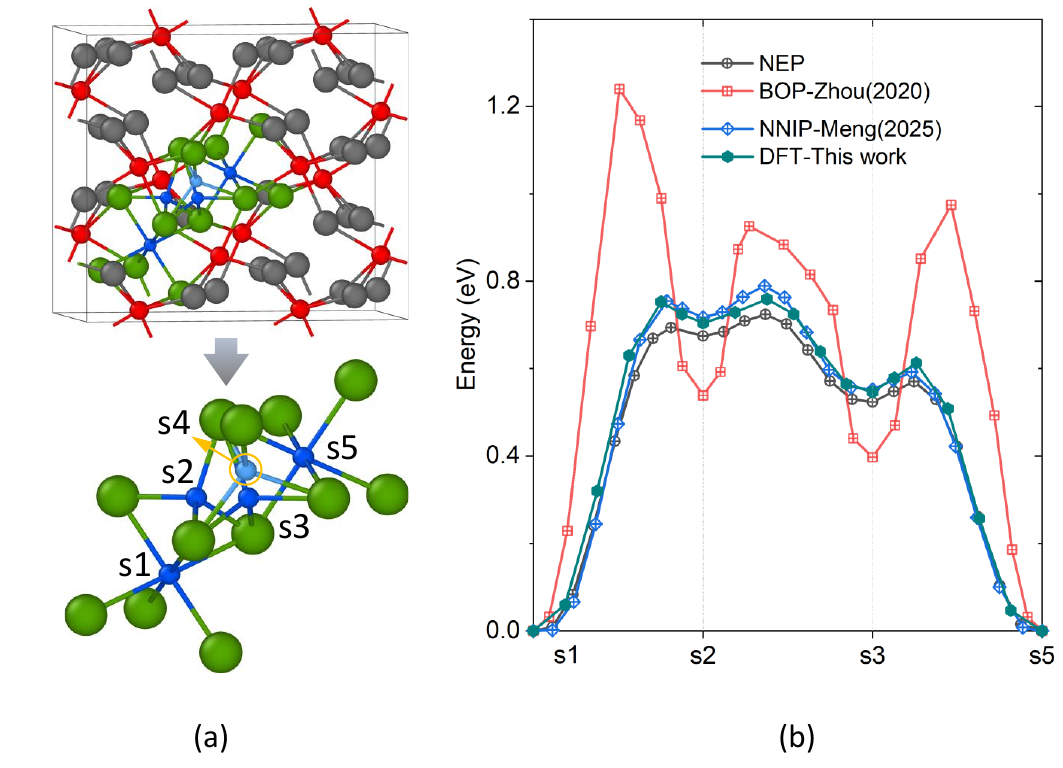}
\caption{\label{HFe3C} Trapping sites along the diffusion path and diffusion barriers of a single H atom in Fe$_3$C. (a) Trapping sites, s1 to s4 indicate the four trapping sites in Fe$_3$C. The local atomic configuration is extracted in the lower panel of (a). Big gray and green,red, and blue balls stand for Fe, C, and H atoms, respectively. (b) Diffusion energy barriers of a single H atom migration from s1 to s4. The results predicted by NNIP\cite{meng2025high}, BOP\cite{zhou2020review}, and DFT are also plotted. }
\end{figure}

There are four independent solution sites for H atom in Fe$_3$C phase, marked as $s1$ to $s4$ in Fig.\ref{HFe3C}(a). Note that the $s5$ is an equivalent site as $s1$.  The H atom exhibits either octahedral or tetrahedral coordination with its neighboring iron atoms at all solution sites.
The most stable solution site is $s1$ with the solution energy of 0.002 eV from the NEP which is in line with -0.02 eV from the NNIP,  0.02 eV from DFT calculation. Comparing to the solution energy of H at bcc lattice of 0.22 eV, $s1$ is energetically preferred trapping site. 
In contrast to $s1$, the solution energies are 0.68, 0.53, and 0.70 eV for $s2$, $s3$, and $s4$, respectively, showing good agreement with the NNIP results (0.69, 0.52, and 0.76 eV), which are closer to the DFT values (0.73, 0.57, and 0.80 eV) than those obtained from the BOP (0.49, 0.35, and 0.43 eV), although the BOP presents the reasonable solution energy.

H diffusion in Fe$_3$C follows an O--T--T--O path, starting from an octahedral site (s1), passing through two tetrahedral sites (s2--s3), and ending at an octahedral site (s5), as depicted in Fig.\ref{HFe3C}(a). The energy profile during the pathway is plotted in Fig.\ref{HFe3C}(b). The results from the NEP are slightly lower than the reference values from DFT calculations but remain within a very small margin of difference. On the BOP side, it can correctly predict the H jumping manner in Fe$_3$C, but overestimates each diffusion energy barrier.

\begin{figure}[H]
    \centering
\includegraphics[width=0.78\textwidth]{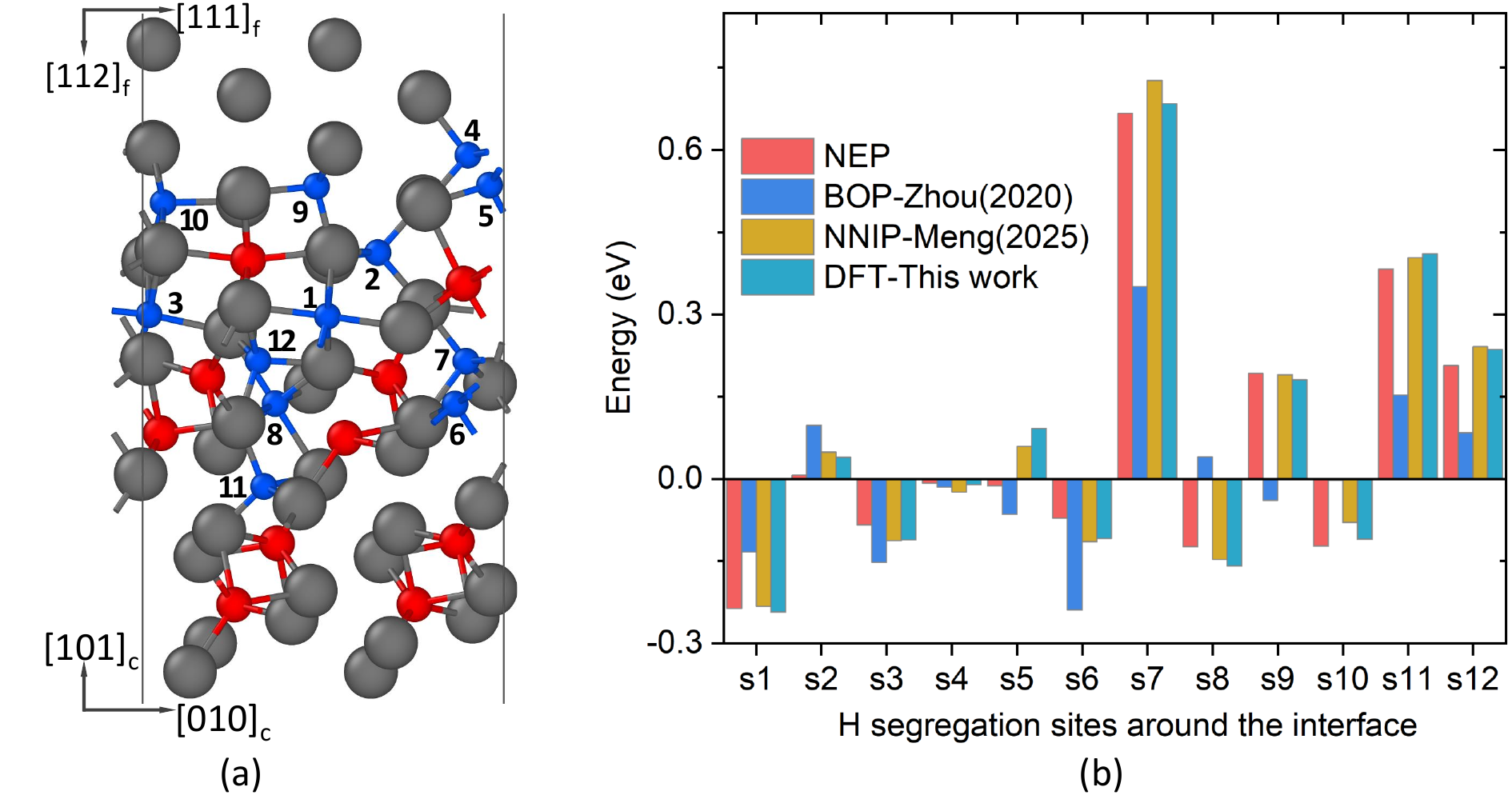}
\caption{\label{HFe3C-Fe} Segregation sites of H atom at a ferrite/cmentite interface and their energies.  (a). Configuration of the ferrite/cementite interface and H segregation sites around the interface. The subscripts $c$ and $f$ denote the orientations of cementite and ferrite, respectively. The gray, red, and blue balls indicate the Fe, C, and H atoms, respectively. (b) Segregation energy 
of H at the sites presented in (a) from the NEP, those predicted by the BOP\cite{zhou2020review}, NNIP\cite{meng2025high}, and DFT are also presented.}
\end{figure}

The pronounced mechanical property of pearlitic steels highly correlates with the ferrite/cementite (Fe/Fe$_3$C) boundaries inside\cite{zhou2017atomic}. The interface configuration with the lowest energy and the corresponding H segregation energies were further analyzed using the NEP and BOP, and compared with those from the NNIP and DFT calculations. The crystallographic orientations of each phase at the interface are shown in Fig. \ref{HFe3C-Fe}(a). The interface energies obtained from NEP and BOP are 0.674 and 0.711 J/m$^2$, respectively. Although both methods overestimate the interface energy compared to the DFT results (0.571 J/m$^2$ \cite{meng2024highly}), the NEP demonstrates higher accuracy than the BOP. 

Fig. \ref{HFe3C-Fe}(b) presents the H segregation energies from the bcc lattice to the twelve interfacial sites as predicted by the NEP, BOP, NNIP, and DFT. The NEP and NNIP predictions show good agreement with the DFT results, while the BOP tends to either overestimate or underestimate the segregation energies. The segregation energy at site $s4$ is nearly zero, indicating a solution energy comparable to that in the bcc lattice, whereas the H segregation energy at site $s8$ is -0.2 eV, similar to the lowest H solution energy in Fe$_3$C ($s1$ in Fig. \ref{HFe3C}(a)).  Of the interfacial sites $s2$, $s5$, $s9$, and $s10$, only $s10$ has a negative segregation energy, yet it is higher than that at $s8$, suggesting a low hydrogen-trapping ability at the interface.

\section{Conclusion}
In this work, we constructed a neuroevolution potential (NEP) for the $\alpha$-Fe--C--H system using a database from a reported work. The computational efficiency was evaluated on both CPU and GPU platforms and compared across different types of potentials, demonstrating that conducting molecular dynamics simulations with NEP on GPUs achieves speed and resource consumption comparable to those of BOP. The NEP exhibits DFT-level accuracy across a wide range of scenarios commonly encountered in studies of $\alpha$-Fe and $\alpha$-Fe–C under hydrogen environments. The NEP shows significant advantages over bond order potentials (BOPs) in describing screw dislocations as well as C/H interactions with screw dislocations. The superior accuracy of the NEP compared to the BOP was further demonstrated for H in cementite and at the ferrite/cementite interface.

Using the NEP on GPUs, atomistic simulations of millions of atoms over nanosecond timescales can be completed within days with DFT-level accuracy, providing a practical approach for investigating hydrogen embrittlement in steel.

\section*{Data availability}
The neuroevolution potential reported in this study will be made available upon reasonable request to the corresponding authors.
\section*{Acknowledgments}

S.O. acknowledges the support by JSPS KAKENHI (Grant Nos. JP23H00161 and JP23K20037), and the supported by MEXT (Ministry of Education, Culture, Sports, Science and Technology of Japan) Programs (Grant Nos. JPMXP1122684766, JPMXP1020230325, and JPMXP1020230327). Part of calculations were performed using computational resources of supercomputer Fugaku provided by the RIKEN Center for Computational Science (Project IDs: hp230205 and hp230212), the large-scale computer systems at the Cybermedia Center, The University of Osaka, and the Large-scale parallel computing server at the Center for Computational Materials Science, Institute for Materials Research, Tohoku University.
																																																						
\section*{CRediT authorship contribution statement}
 \textbf{Fan-Shun Meng:} Conceptualization, Methodology, Investigation, Validation, Writing – original draft.  
\textbf{Shuhei Shinzato:} Conceptualization, Methodology, Investigation, Validation, Writing – original draft. 
\textbf{Zhiqiang Zhao:} Conceptualization, Methodology. 
\textbf{Junping Du:} Methodology, Investigation, 
\textbf{Lei Gao:} Methodology, 
\textbf{Zheyong Fan:}  Writing – review $\&$ editing.
\textbf{Shigenobu Ogata:} Resources, Conceptualization, Methodology, Writing--review $\&$ editing, Supervision, Funding acquisition.

\section*{Competing interests}
The authors declare that they have no known competing interests.
\bibliographystyle{elsarticle-num}

\end{document}